\documentclass{article}

 \PassOptionsToPackage{numbers, compress}{natbib}

 \usepackage[preprint]{nips_2018}

\usepackage[utf8]{inputenc} 
\usepackage[T1]{fontenc}    
\usepackage{hyperref}       
\usepackage{url}            
\usepackage{booktabs}       
\usepackage{amsfonts}       
\usepackage{nicefrac}       
\usepackage{microtype}      

\usepackage[utf8]{inputenc} 
\usepackage[english]{babel} 

\usepackage{graphicx}
\usepackage{mathtools, cuted}
\usepackage{hyperref}
\usepackage{amsmath}
\usepackage{amssymb}
\usepackage{mathtools}
\usepackage{amsthm}

\newtheorem{theorem}{Theorem}[section]
\newtheorem{definition}[theorem]{Definition}

\newtheorem{lemma}[theorem]{Lemma}

\title{CM Sequence based Trajectory Modeling with Destination}

\author{
  Reza Rezaie and X. Rong Li \\
 Department of Electrical Engineering\\
 University of New Orleans\\
New Orleans, LA 70148 \\
  \texttt{rrezaie@uno.edu} and \texttt{xli@uno.edu} \\
}

\begin{document}

\maketitle

\begin{abstract}
In some problems there is information about the destination of a moving object. An example is an airliner flying from an origin to a destination. Such problems have three main components: an origin, a destination, and motion in between. To emphasize that the motion trajectories end up at the destination, we call them \textit{destination-directed trajectories}. The Markov sequence is not flexible enough to model such trajectories. Given an initial density and an evolution law, the future of a Markov sequence is determined probabilistically. One class of conditionally Markov (CM) sequences, called the $CM_L$ sequence (including the Markov sequence as a special case), has the following main components: a joint endpoint density (i.e., an initial density and a final density conditioned on the initial) and a Markov-like evolution law. This paper proposes using the $CM_L$ sequence for modeling destination-directed trajectories. It is demonstrated how the $CM_L$ sequence enjoys several desirable properties for destination-directed trajectory modeling. Some simulations of trajectory modeling and prediction are presented for illustration.

\end{abstract}


\section{Introduction}

Markov processes have been widely used in many applications. A Markov process has two main components: an initial density and an evolution law. Consider the problem of trajectory modeling and prediction with destination information, e.g., an airliner's trajectory from an origin to a destination. Such a problem has three main components: an origin, motion, and a destination, for which a Markov process does not fit since it can not take the destination information into account. In other words, the destination density of a Markov process is completely determined by its initial density and evolution law. So, a more general process is needed for modeling destination-directed trajectories.

Trajectory modeling and prediction with an intent or a destination has been studied in the literature. Some intent-based trajectory prediction approaches for air traffic control (ATC) can be found in \cite{Hwang0}--\cite{Krozel2}. \cite{Hwang0}--\cite{Hwang2} presented some trajectory prediction approaches based on hybrid estimation aided by intent information. Also, in \cite{Louis} the interacting multiple model (IMM) approach was used for trajectory prediction, where a higher weight was assigned to the model with the closest heading towards the waypoint. \cite{Krozel} presented an approach for trajectory prediction using an inferred intent based on a database. \cite{Krozel2} discussed the use of waypoint information for trajectory prediction in ATC. Ship trajectories were modeled by a Gauss-Markov model in \cite{Ship}, where predictive information was incorporated. In \cite{BV_Krener}, a boundary value linear system was used for modeling trajectories with a specific destination. After quantizing the state space, \cite{Fanas1}--\cite{White_Tracking1} used finite-state reciprocal sequences for intent inference and trajectory modeling. A problem with quantized state space is the complexity of the corresponding estimation algorithms. So, the complexity of the algorithms used in \cite{Fanas1}--\cite{White_Tracking1} was also addressed. \cite{S_1}--\cite{S_2} used bridging distributions for the purpose of intent inference, for example, in selecting an icon on an in-vehicle interactive display. A general class of stochastic sequences that naturally models destination-directed trajectories is missing in the literature. The goal of this paper is to start from the main components of destination-directed trajectories and develop a framework and a class of stochastic sequences that naturally models such trajectories.

Inspired by \cite{Mehr}, Gaussian CM sequences were studied in \cite{CM_Part_I_Conf}--\cite{CM_Explicitly}. There are several classes of CM sequences, one of which is called $CM_L$ \cite{CM_Part_I_Conf}. A sequence defined over the interval $[0,N]$ is $CM_L$ if and only if (iff) conditioned on the state at time $N$, the sequence is Markov over $[0,N-1]$. The subscript ``$L$" is used because the conditioning is at the \textit{last} time of the interval. A dynamic model (called $CM_L$ model) governing the nonsingular Gaussian (NG) $CM_L$ sequence was presented in \cite{CM_Part_I_Conf}. Reciprocal processes were introduced in \cite{Bernstein} connected to a problem posed by E. Schrodinger \cite{Schrodinger_1}--\cite{Schrodinger_2}. Later, reciprocal processes were studied more in \cite{Slepian}--\cite{Moura}. Every reciprocal sequence is a $CM_L$ sequence \cite{CM_Part_II_A_Conf}. A dynamic model of the NG reciprocal sequence was presented in \cite{Levy_Dynamic}, which shows that the evolution of the NG reciprocal sequence can be described by a second-order nearest-neighbor model driven by a locally correlated dynamic noise. This model can be seen as a generalization of the Markov model. However, due to the nonwhiteness of the dynamic noise and the nearest-neighbor structure, it is not necessarily easy to apply. That is why different approaches have been presented for recursive estimation of reciprocal sequences based on that model \cite{Levy_Dynamic}--\cite{Moura}. In \cite{CM_Part_II_A_Conf} the reciprocal sequence was studied from the CM viewpoint and (based on the $CM_L$ model) a model, called a reciprocal $CM_L$ model, was presented for the evolution of the NG reciprocal sequence. This model is driven by white dynamic noise and is simple and easily applicable.

The main elements of the $CM_L$ sequence are a joint endpoint density (in other words, an initial density and a final density conditioned on the initial) and an evolution law. Having a Markov-like property, the evolution law of the $CM_L$ sequence is simple (and thus easily applicable) and conceptually appealing for modeling destination-directed trajectories. The $CM_L$ sequence can model well the main elements of destination-directed trajectories. 

The main contribution of this paper is that it proposes the use of the $CM_L$ sequence for destination-directed trajectory modeling. It demonstrates that the Markov sequence is not flexible for modeling destination-directed trajectories. Then, it shows how the $CM_L$ sequence can naturally model such trajectories. Also, it discusses that the $CM_L$ sequence enjoys desirable properties for modeling destination-directed trajectories. Simulations for destination-directed trajectory modeling and prediction are presented. This paper builds upon our previous results about CM and reciprocal sequences \cite{CM_Part_I_Conf}--\cite{CM_Part_II_B_Conf} to show their application in trajectory modeling.
  
In Section \ref{DD_Modeling}, the $CM_L$ sequence and its model are presented for trajectory modeling. In Section \ref{CML_Parameter}, the $CM_L$ model parameter design is discussed. Simulations are presented in Section \ref{Simulation}. In Section \ref{Summary}, desirable properties of the $CM_L$ sequence for destination-directed trajectory modeling and prediction are discussed.

\section{Destination-Directed Trajectory Modeling Using $CM_L$ Sequence}\label{DD_Modeling}

We consider the following notation for index (time) interval and stochastic sequences: $[x_k]_{i}^{j} \triangleq \lbrace x_k, k \in [i,j] \rbrace , i<j$, and $[x_k] \triangleq [x_k]_{0}^{N}$, where $[i,j] \triangleq \lbrace i,i+1,\cdots ,j-1,j \rbrace$. Also, ZMNG and NG stands for ``zero-mean nonsingular Gaussian" and "nonsingular Gaussian", respectively. The symbol ``$\setminus$" is used for set subtraction. $F(\cdot|\cdot)$ is a conditional cumulative distribution function (CDF). $p(\cdot)$ and $p(\cdot|\cdot)$ denote probability density function (PDF) and conditional PDF.

\begin{definition}\label{Markov} 
$[x_k]$ is Markov if $F(\xi_k|[x_i]_0^j) =F(\xi_k|x_{j})$, $\forall j,k \in [0,N], j<k$, $\forall \xi _k \in \mathbb{R}^d$, where $d$ is the dimension of $x_k$.

\end{definition}

To model the trajectory of a moving object without considering its destination, there are two main components: the origin and the evolution. Accordingly, a Markov sequence is determined by two components: an initial density and an evolution law. Sample paths of a Markov sequence can be used for modeling such trajectories. For example, a nearly constant velocity/acceleration/turn (with white noise) motion model is a Markov model. A dynamic model for the evolution of the Gaussian Markov sequence is as follows.
\begin{lemma}\label{Model_Dynamic_Proposition}
A ZMNG $[x_k]$ is Markov iff
\begin{align}
x_k=M_{k,k-1}x_{k-1}+e_{k}, k \in [1,N], \quad x_0=e_0 \label{Markov_Model}
\end{align}
and $[e_k]$ ($M_k=\text{Cov}(e_k)$) is a zero-mean white NG sequence.
\end{lemma}
     
Markov sequences are not flexible enough for modeling an origin, evolution, and a destination. This is because the destination density of a Markov sequence is determined by its initial density and evolution law. Therefore, a more general class of stochastic sequences is needed. 

Let destination-directed trajectories be modeled as the sample paths of a sequence $[x_k]$. In probability theory, one can interpret the main elements of destination-directed trajectories as follows. The origin (destination) is modeled by a density function of $x_0$ ($x_N$). The relationship between the origin and the destination is modeled by their joint density, i.e., joint density of $x_0$ and $x_N$. Since the destination of the trajectories (i.e., density of $x_N$) is known, the evolution law can be modeled as a conditional density (over the space of sample paths) given the state at destination $x_N$. Different choices of this conditional density correspond to different evolution laws. The simplest choice is conditioned on $x_N$ the density being equal to the product of its marginals: $p([x_k]_0^{N-1}|x_N)=\prod _{k=0}^{N-1}p(x_k|x_N)$. However, this choice of the conditional density is often too simple to be suitable. Then, the next step is to choose the conditional density corresponding to the Markov sequence: $p([x_k]_0^{N-1}|x_N)=p(x_0|x_N)\prod _{k=1}^{N-1}p(x_k|x_{k-1},x_N)$. This evolution law corresponds to the $CM_L$ sequence. The main elements of a $CM_L$ sequence $[x_k]$ are: a joint density of $x_0$ and $x_N$ (in other words, an initial density and a final density conditioned on the initial, or equivalently, the other way round) and an evolution law, where the evolution law is conditionally Markov (conditioned on $x_N$). The above argument naturally leads to $CM_L$ sequences for modeling destination-directed trajectories. Following the same argument, we can consider more general and complicated evolution laws, if necessary. For example, the conditional law (density) (conditioned on $x_N$) can be higher order Markov instead of the first order Markov. Therefore, by choosing the conditional law, all destination-directed trajectory models can be obtained. 

\begin{definition}\label{CML}
$[x_k]$ is $CM_L$ if $F(x_k|[x_i]_0^j,x_N) =F(x_k|$ $x_j,x_N)$, $\forall j, k \in [0,N-1], j<k$, $\forall \xi _k \in \mathbb{R}^d$, where $d$ is the dimension of $x_k$.

\end{definition}

In other words, $[x_k]$ is $CM_L$ iff $[x_k]_{j+1}^{N-1}$ and $[x_k]_{0}^{j-1}$ are independent conditioned on $x_{j}$ and $x_{N}$ ($\forall j \in [1,N-2]$). A $CM_L$ model is as follows \cite{CM_Part_I_Conf}. 
\begin{theorem}\label{CML_Forward_Dynamic_Proposition}
A ZMNG $[x_k]$ is $CM_L$ iff
\begin{align}
x_k&=G_{k,k-1}x_{k-1} + G_{k,N}x_N+e_k, k \in [1,N-1] \label{CML_Dynamic}\\
x_N&=e_N, \quad x_0=G_{0,N}x_N+e_0\label{BCondition_2T}
\end{align}
and $[e_k]$ ($G_k=\text{Cov}(e_k)$) is a zero-mean white NG sequence.

\end{theorem}

For trajectory modeling we need non-zero-mean sequences. A non-zero-mean NG sequence $[x_k]$ is $CM_L$ (Markov) iff its zero-mean part is governed by $\eqref{CML_Dynamic}$--$\eqref{BCondition_2T}$ ($\eqref{Markov_Model}$). For simplicity, we present the results for the zero-mean case; however, in the simulation non-zero-mean sequences are considered. The $CM_L$ model governing the non-zero-mean NG $CM_L$ sequences considered in the simulation is as follows. Let $\mu _0$ ($\mu _N$) and $C_0$ ($C_N$) be the mean and covariance of the origin (destination) density. Also, let $C_{0,N}$ be the cross-covariance of the states at the origin and the destination. We have $x_N \sim \mathcal{N}(\mu _N, C_N)$. Then, $x_0=\mu _0+G_{0,N}(x_N-\mu _N)+e_0$, where $G_{0,N}=C_{0,N}C_N^{-1}$, $G_0=C_0-C_{0,N}C_N^{-1}(C_{0,N})'$, and $G_N=C_N$. Then, the state evolution for $k \in [1,N-1]$ is governed by $\eqref{CML_Dynamic}$.

In $\eqref{CML_Dynamic}$, $x_N$ is generated first and then followed by the other states. Therefore, the model is not causal. However, we should notice that for estimation (tracking/prediction), the non-causality of model $\eqref{CML_Dynamic}$ requires \textit{information} about $x_N$ (i.e., $p(x_N)$), which is available. Therefore, this model is totally applicable.

\section{$CM_L$ Model Parameter Design for Destination-Directed Trajectory Modeling}\label{CML_Parameter}

To use the $CM_L$ model for the evolution of destination-directed trajectories, we need an approach for design of its parameters. In the following, such an approach is presented \cite{CM_Part_II_B_Conf}. Before doing so, we review the reciprocal sequence as a special case of the $CM_L$ sequence.

\begin{lemma}\label{CDF}
$[x_k]$ is reciprocal iff $F(\xi _k|[x_{i}]_{0}^{j},[x_i]_l^N)=F(\xi _k|x_j,x_l)$, $\forall j,k,l \in [0,N]$ ($j < k < l$), $\forall \xi _k \in \mathbb{R}^d$, where $d$ is the dimension of $x_k$.  

\end{lemma}

Theorem \ref{CML_R_Dynamic_Forward_Proposition} below presents the conditions under which the $CM_L$ model governs a reciprocal sequence \cite{CM_Part_II_A_Conf}. Such a $CM_L$ model is called a \textit{reciprocal} $CM_L$ model.

\begin{theorem}\label{CML_R_Dynamic_Forward_Proposition}
A ZMNG $[x_k]$ is reciprocal iff it is governed by $\eqref{CML_Dynamic}$--$\eqref{BCondition_2T}$ and $G_k^{-1}G_{k,N}=G_{k+1,k}'G_{k+1}^{-1}G_{k+1,N}, k \in [1,N-2]$. Moreover, $[x_k]$ is Markov iff additionally we have $G_0^{-1}G_{0,N}=G_{1,0}'G_1^{-1}G_{1,N}$.

\end{theorem}

An approach for the $CM_L$ model parameter design for modeling destination-directed trajectories is as follows. Such trajectories can be modeled by combining (superimposition of) two key assumptions: (i) the moving object follows a Markov model $\eqref{Markov_Model}$ (e.g., a nearly constant velocity model) without considering the destination information, and (ii) the destination density is known (which can differ from the destination density of the Markov model in (i)). Note that for theoretical purposes we assume the destination density is known. But in a real problem an approximate destination density can be used. The above two assumptions are valid in a real problem of trajectory modeling with destination. Let $[s_k]$ be a Markov sequence governed by $\eqref{Markov_Model}$ (e.g., a nearly constant velocity model). Since every Markov sequence is $CM_L$, $[s_k]$ can also obey a $CM_L$ model as
 \begin{align}
s_k&=G_{k,k-1}s_{k-1}+G_{k,N}s_N+e^s_k, k \in [1,N-1]\label{CML_Dynamic_for_Markov}\\
s_N&=e^{s}_N, \quad s_0=G^{s}_{0,N}s_N+e^{s}_0\label{CML_R_FQ_BC2}
\end{align}
where $[e^s_k]$ is a zero-mean white NG sequence with covariances $G_k, k \in [1,N-1]$, $G^s_0$, and $G^s_N$.

Parameters of $\eqref{CML_Dynamic_for_Markov}$ can be obtained as follows. By $\eqref{Markov_Model}$, we have $p(s_k|s_{k-1})=\mathcal{N}(s_k;M_{k,k-1}s_{k-1},M_{k})$. Since $[s_k]$ is Markov, we have ($ k \in [1,N-1]$)
\begin{align}
p(s_k|s_{k-1},s_N)&=\frac{p(s_k|s_{k-1})p(s_N|s_k,s_{k-1})}{p(s_N|s_{k-1})}\nonumber\\
&=\frac{p(s_k|s_{k-1})p(s_N|s_k)}{p(s_N|s_{k-1})}\label{CML_Reciprocal_Transition}\\
&=\mathcal{N}(s_k;G_{k,k-1}s_{k-1}+G_{k,N}s_N;G_k)\nonumber
\end{align}
where $G_{k,k-1}$, $G_{k,N}$, and $G_k$ are obtained as
\begin{align}
G_{k,k-1}&=M_{k,k-1}-G_{k,N}M_{N|k-1} \label{CML_Choice_1}\\
G_{k,N}&=G_kM_{N|k}'C_{N|k}^{-1} \label{CML_Choice_2}\\
G_k&=(M_{k}^{-1}+M_{N|k}'C_{N|k}^{-1}M_{N|k})^{-1}\label{CML_Choice_3}\\
M_{N|k}&=M_{N,N-1}\cdots M_{k+1,k}, k \in [1,N-1], M_{N|N}=I\nonumber\\
C_{N|k}&=\sum _{n=k}^{N-1}M_{N|n+1}M_{n+1}M_{N|n+1}', k \in [1,N-1]\nonumber
\end{align}
\begin{align*}
p(s_N|s_i)&=\mathcal{N}(s_N;M_{N|i}s_{i},C_{N|i}), i \in [0,N-1]\nonumber
\end{align*}

Now, we construct a sequence $[x_k]$ governed by
\begin{align}
x_k&=G_{k,k-1}x_{k-1}+G_{k,N}x_N+e_k, k \in [1,N-1]\label{CML_Dynamic_for_Markov_x}\\
x_N&=e_N, \quad x_0=G_{0,N}x_N+e_0\label{CML_R_FQ_BC2_x}
\end{align}
where $[e_k]$ is a zero-mean white NG sequence with covariances $G_k, k \in [1,N-1], G_0$, and $G_N$. Note that $\eqref{CML_Dynamic_for_Markov_x}$ and $\eqref{CML_Dynamic_for_Markov}$ have the same parameters ($G_{k,k-1}, G_{k,N},G_k, k \in [1,N-1]$), but parameters of $\eqref{CML_R_FQ_BC2_x}$ ($G_{0,N},G_0,G_N$) and parameters of $\eqref{CML_R_FQ_BC2}$ ($G^s_{0,N},G^s_0,G^s_N$) are different. Parameters of $\eqref{CML_R_FQ_BC2_x}$ ($G_{0,N},G_0,G_N$) can be chosen arbitrarily (i.e. $G_{0,N}$ can be any matrix with suitable dimension, and $G_0$ and $G_N$ any positive definite matrix with suitable dimension). Thus, $[x_k]$ can have any joint density of $x_0$ and $x_N$. So, $[s_k]$ and $[x_k]$ have the same $CM_L$ model ($\eqref{CML_Dynamic_for_Markov}$ and $\eqref{CML_Dynamic_for_Markov_x}$) (in other words, the same transition density $\eqref{CML_Reciprocal_Transition}$), but $[x_k]$ can have any joint endpoint density. In other words, any origin and destination of $[x_k]$ can be so modeled. By Theorem \ref{CML_Forward_Dynamic_Proposition}, $[x_k]$ is a $CM_L$ sequence. Therefore, combining assumptions (i) and (ii) above naturally leads to the $CM_L$ sequence $[x_k]$ whose $CM_L$ model is the same as that of $[s_k]$ while the former can model any origin and destination. Thus, model $\eqref{CML_Dynamic_for_Markov_x}$ with $\eqref{CML_Choice_1}$--$\eqref{CML_Choice_3}$ is a desirable model for destination-directed trajectory modeling based on (i) and (ii) above. Such a $CM_L$ model is used in Section \ref{Simulation}. Since it is obtained from a Markov model, $\eqref{CML_Dynamic_for_Markov_x}$ is called a $CM_L$ model \textit{induced} by a Markov model.

The following theorem shows that an induced $CM_L$ model (i.e., $\eqref{CML_Dynamic_for_Markov_x}$ with $\eqref{CML_Choice_1}$--$\eqref{CML_Choice_3}$) is actually a reciprocal $CM_L$ model. So, all its governed sequences $[x_k]$ are reciprocal (for any choice of parameters of boundary condition $\eqref{CML_R_FQ_BC2_x}$). Also, it shows that every reciprocal $CM_L$ model can be induced by a Markov model following the above approach \cite{CM_Part_II_B_Conf}.

\begin{theorem}\label{CML_R_Dynamic_FQ_Proposition} 
A ZMNG $[x_k]$ is reciprocal iff it obeys $\eqref{CML_Dynamic}$--$\eqref{BCondition_2T}$, where $(G_{k,k-1},G_{k,N},G_k)$, $k \in [1,N-1]$, are given by $\eqref{CML_Choice_1}$--$\eqref{CML_Choice_3}$, $M_{k,k-1}$, $ k \in [1,N]$, are square matrices, and $M_k$, $k \in [1,N]$, are positive definite matrices with the dimension of $x_k$.

\end{theorem}

The idea of obtaining a reciprocal evolution law from a Markov evolution law was used in \cite{Schrodinger_1}, \cite{Jamison_Reciprocal}, and later for finite-state reciprocal sequences in \cite{Fanas1}, \cite{White_3}. However, first, our reciprocal $CM_L$ model induced as above (Theorem \ref{CML_R_Dynamic_FQ_Proposition}) is from the viewpoint of the CM sequence. Second, Theorem \ref{CML_R_Dynamic_FQ_Proposition} shows that \textit{every} reciprocal $CM_L$ model can be induced by a Markov model (i.e, necessity and sufficiency). 

The above approach is only one method for the $CM_L$ model parameter design for trajectory modeling. A more general approach for the $CM_L$ model parameter design was sketched in \cite{CM_Part_II_B_Conf}, based on a representation of a $CM_L$ sequence in terms of a Markov sequence.

\section{Simulation}\label{Simulation}

Destination-directed trajectories are modeled by a NG $CM_L$ sequence. Model $\eqref{CML_Dynamic}$ can be written as ($k \in [1,N-1]$)
\begin{align}
\left[\begin{array}{c}
x_k\\
x_N
\end{array}\right]&=\left[ \begin{array}{cc}
G_{k,k-1} & G_{k,N}\\
0 & I
\end{array}\right]\left[\begin{array}{c}
x_k\\
x_N
\end{array}\right]+\left[\begin{array}{c}
e_k\\
0
\end{array}\right]
\end{align}
Also, the measurement equation is $z_k=Hx_k+v_k, k \in [1,N], H=\left[\begin{array}{cccc}
1 & 0 & 0 & 0\\
0 & 0 & 1 & 0
\end{array}\right]$, where $[v_k]_1^{N}$ ($\text{Cov}(v_k)=\text{diag}(100,100)$) is a zero-mean white NG sequence uncorrelated with $[x_k]$. Results from linear system theory can be used for trajectory filtering and prediction. We skip them.

\begin{figure}
\centering
    \includegraphics[scale=0.42]{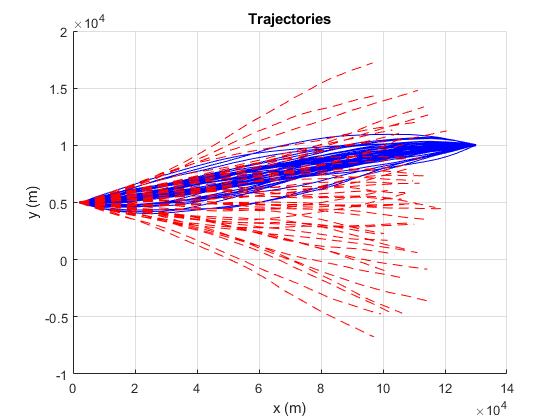}
\caption{$CM_L$ (solid lines) vs. Markov (dash lines) model for trajectory modeling with destination.}
\label{F1}
\end{figure}

\begin{figure}
\centering
    \includegraphics[scale=0.38]{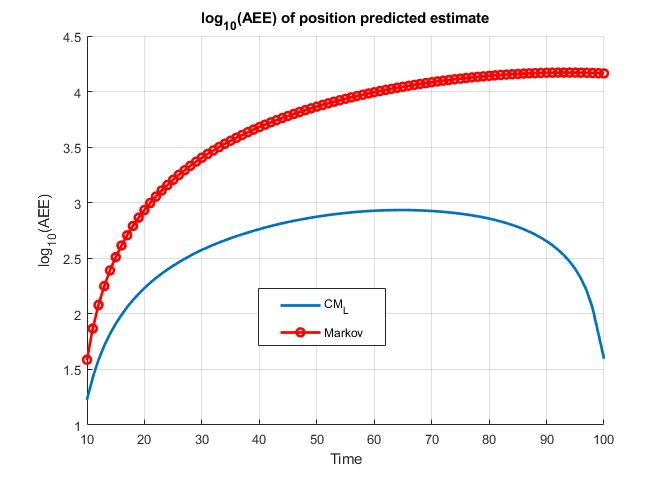}
\caption{Logarithm of AEE of position prediction ($\text{log}_{10}(\text{AEE}^p_{9+n|9})$).}
\label{F2}
\end{figure}

Consider a two-dimensional scenario, where the state of a moving object at time $k$ is $x_k=[ \mathsf{x} , \mathsf{\dot{x}}, \mathsf{y} , \mathsf{\dot{y}}]_k'$ with position $[\mathsf{x},\mathsf{y}]'$ and velocity $[\mathsf{\dot{x}},\mathsf{\dot{y}}]'$. The origin and destination densities are assumed to be Gaussian. Means and covariances of the origin ($\mu _0$ and $C_0$) and destination ($\mu _N$ and $C_N$) densities are $\mu _0=[ 2000 , 70 , 5000 , 0 ]'$, $C_0=\text{diag}(1000,10,1000,10)$, $\mu _N=[ 130000 , 70 , 10000 , 0 ]'$, and $C_N=\text{diag}(1000,10,1000,10)$. Only for simplicity the origin and the destination are assumed to be statistically independent (i.e., $G_{0,N}=0$ in $\eqref{BCondition_2T}$). In fact, the $CM_L$ modeling is much more powerful if $G_{0,N}\neq 0$ in $\eqref{BCondition_2T}$ (to model the relationship between the origin and the destination). We consider a $CM_L$ model induced by a Markov model (Theorem \ref{CML_R_Dynamic_FQ_Proposition}). Parameters of the corresponding Markov model $\eqref{Markov_Model}$ are $M_{k,k-1}=\text{diag}(F,F), M_{k}=\text{diag}(Q,Q), k \in [1,N]$, where $F=\left [\begin{array}{cc}
1 & T\\
0 & 1
\end{array}\right]$, $Q=q\left[
\begin{array}{cc}
T^3/3 & T^2/2\\
T^2/2 & T
\end{array} \right]$, $T=15$ second, $q=0.01$, and $N=100$. Fig. \ref{F1} compares the Markov model and the $CM_L$ model for trajectory modeling (the 50 solid lines are trajectories of the $CM_L$ sequence and the 50 dash lines are trajectories of the Markov sequence). Both sequences model the origin well. Also, near the origin the two models differ a little. However, later their difference grows. This is due to the poor performance of the Markov model in incorporating the destination information. Fig. \ref{F2} shows the log of the average Euclidean errors (AEE) \cite{Li_AEE} of the predicted estimates of the position obtained based on the $CM_L$ and the Markov model. The AEE of the position predicted estimate ($\text{AEE}^{p}_{k+n|k}$) is $\frac{1}{M}\sum _{i=1}^{M}\sqrt{(\mathsf{x}_{k+n}-\hat{\mathsf{x}}_{k+n|k})^2+(\mathsf{y}_{k+n}-\hat{\mathsf{y}}_{k+n|k})^2}$, where $[\mathsf{x}_{k+n} , \mathsf{y}_{k+n}]'$ is the true position at $k+n$ ($k+n=10, \ldots , 100$) and $[\hat{\mathsf{x}}_{k+n|k} , \hat{\mathsf{y}}_{k+n|k}]'$ is its prediction using measurements up to time $k=9$, and $M=1000$ is the number of Monte Carlo runs. The ratio of $\text{AEE}^p_{100|9}$ of the Markov model over $\text{AEE}^p_{100|9}$ of the $CM_L$ model $\frac{\text{AEE}^p_{100|9}(\text{Markov})}{\text{AEE}^p_{100|9}(CM_L)}$ is $368.13$, which is huge. 

\section{Conclusions}\label{Summary}

We call motion trajectories that end up at a specific destination, \textit{destination-directed trajectories}. The main components of these trajectories are: an origin, a destination, and evolution. The main components of the $CM_L$ sequence are a joint endpoint density (i.e., an initial density and conditioned on it a destination density) and a Markov-like evolution law. The $CM_L$ sequence is conceptually appealing, flexible, and simple to apply for modeling destination-directed trajectories. First, the $CM_L$ sequence is a good fit for destination-directed trajectory modeling and prediction. Second, it outperforms Markov-based trajectory models very much. Third, the $CM_L$ sequence is a powerful tool, capable of modeling different scenarios. For example, assume that according to the air traffic control near an airport, airliners are supposed to enter the terminal area from some prespecified direction. This information can be used to specify a destination density of a $CM_L$ sequence. Fourth, the $CM_L$ sequence can model the correlation between the states of an airliner at the origin and the destination airports. Fifth, there is no restriction on the parameters of the $CM_L$ dynamic model, which makes its analysis easily possible. Also, since there is no restriction on the parameters, they can be easily determined, e.g., based on an optimality criterion for trajectory design. It is possible based on a representation of a $CM_L$ sequence in terms of a Markov sequence presented in \cite{CM_Part_II_B_Conf}. Sixth, the $CM_L$ dynamic model provides a natural way for the update (uncertainty reduction) of the available information about the destination of an airliner as more measurements are received. In addition, CM sequences provide a simple framework for trajectory modeling in different scenarios, e.g., a CM sequence was proposed in \cite{DW_Conf} for trajectory modeling with waypoint information.

\subsubsection*{Acknowledgments}

Research is partly supported by NASA/LEQSF Phase03-06 through grant NNX13AD29A.

\end{document}